# Effect of shallow traps on polaron transport at the surface of organic semiconductors

M. F. Calhoun, C. Hsieh and V. Podzorov[*]

*Department of Physics and Astronomy, Rutgers University, Piscataway, New Jersey, USA*

The photo-induced electron and hole transfer across the semiconductor-dielectric interface in trap dominated *p*-type organic field-effect transistors (OFETs) has been investigated. It has been observed that the transfer of electrons into the dielectric results in a decrease of the field-effect mobility of polarons, suggesting that additional shallow traps are generated in the conduction channel. Using this effect, the dependence of the field-effect mobility on the density of shallow traps, $\mu(N)$, has been measured, which allowed to estimate the average polaron trapping time, $\tau_{tr} = 50 \pm 10$ ps, and the density of shallow traps, $N_0 = (3 \pm 0.5)\cdot 10^{11}$ cm$^{-2}$, in the channel of single-crystal tetracene devices.

[*]Electronic mail: podzorov@physics.rutgers.edu

Charge transport in organic semiconductors is due to the motion of polarons that readily interact with defects in the form of charge trapping [1]. The recent development of single-crystal organic field-effect transistors (OFETs) (see, e.g., [2]) with significantly reduced disorder enabled the realization of intrinsic (not dominated by trapping) polaronic transport at organic surface [3,4,5]. Nevertheless, the majority of OFETs still operates in the trap dominated regime, in which the average trapping time of polarons at shallow traps, $\tau_{tr}$, is much longer than the average time of diffusive motion between the trapping events, $\tau$ ($\tau_{tr} >> \tau$). For this reason, studies of shallow traps in organic semiconductors are very important. Traditional methods of defect characterization in semiconductors, such as space charge limited current [6,7], time-of-flight [6] and photo-current [8] spectroscopies, are useful for probing localized states in the bulk. However, assessing the electronic properties of surfaces and interfaces would be especially valuable both for applications [9] and for the fundamental understanding of surface polarons [10]. While measurements of the field effect threshold voltage ($V_{th}$) allow determination of the interfacial density of deep traps [3,11], shallow traps that limit the charge carrier mobility ($\mu$) can not be characterized in a similar way. Traps at the interface can be studied by capacitance measurements in metal-insulator-semiconductor structures [12] – this technique, however, cannot directly relate the trap parameters with the conduction along the interfaces.

Here, we describe an effect that can be used to extract information on shallow traps in the conduction channel of OFETs. It is based on photo-induced charge transfer across the interface - a process responsible for a positive or negative shift of the onset voltage ($\Delta V_{onset}$) without changes of $\mu$ in rubrene OFETs operating in the intrinsic regime [13]. The same process in trap-dominated devices results in a different response: $\mu$ monotonically decreases with the amount of electrons transferred into the dielectric; the transfer of holes does not change $\mu$. Our interpretation of this effect is based on an assumption that electrons immobilized in the dielectric near the interface create local potential "wells" that act as effective polaron shallow traps, whereas transferred holes produce potential "bumps" that might scatter polarons, but cannot trap them. This effect provides the means to vary the density of shallow traps, $N$, in an OFET's channel and to measure the dependence $\mu(N)$. In the framework of Multiple Trap and Release (MTR) model [14], such a dependence can be used to extract the average trapping time, $\tau_{tr}$, and the initial density of shallow traps in the channel, $N_0$.

Free-standing single-crystal *tetracene* OFETs with parylene gate dielectric and a semitransparent gate (120 Å-thick Ag film) have been used (for the fabrication details see, e.g., [2]). It has been verified that the devices are not contact-limited in the range of investigated temperatures [15]. The central result of this study is presented in Fig. 1. The transconductance characteristics, $I_{SD}(V_g)$, have been measured in the dark, after the channel of the device has been illuminated through the gate at different applied gate voltages ($V_g^{illum}$). The illumination technique is described elsewhere [13]. Photons absorbed in tetracene generate charge carriers near the interface, and depending on the sign of $V_g^{illum}$, either electrons or holes are transferred and immobilized in the disordered insulator [13]. At a fixed $V_g^{illum}$ this process saturates with time; therefore, illuminations at different $V_g^{illum}$ have been used to obtain different densities of the transferred charge (Fig. 1). The resulting positive or negative shift of $V_{onset}$ can be accurately measured by recording $I_{SD}(V_g)$ curves in the dark and plotting them on a semi-log scale (in the insets in Fig. 1, $V_{onset}$ is shown by the arrow).

It can be seen that in addition to the shift of $V_{onset}$, the electron transfer results in a monotonic decrease of the slope of $I_{SD}(V_g)$, indicating a decrease of the mobility of polarons in the channel. The effect is very persistent: $\mu$ does not recover to the initial value with time. The density of the transferred charge can be calculated as $\Delta N = C_i \cdot \Delta V_{onset}/e$, where $C_i$ is the gate-channel capacitance per unit area, and $e$ is the elementary charge. The dependence of the mobility on the density of transferred charge, $\mu(\Delta N)$, extracted from Fig. 1, is shown in Fig. 2. The density of deep traps, defined as $N_{deep} = C_i \cdot (V_{th} - V_{onset})/e$, where $V_{th}$ is the voltage at which the linear extrapolation of $I_{SD}(V_g)$ intercepts $V_g$-axis, is also shown in Fig. 2 for the electron transfer. It is evident that the transferred electrons do not change $N_{deep}$. Below we show that in the model of *diffusive motion of polarons in the presence of shallow traps*, the experimental dependence $\mu(\Delta N)$ can be used to estimate the density of shallow traps in the field-induced channel, $N_0$, and the average trapping time, $\tau_{tr}$.

According to the MTR model, polarons diffusively move with an intrinsic mobility $\mu_0$ between the trapping events that reduce $\mu_0$ to an effective value, $\mu$:

$$\mu = \mu_0 \frac{\tau}{\tau + \tau_{tr}}. \tag{1}$$

Among all the in-gap states, *shallow traps* are the most important in the MTR model. They occupy the energy window of the order of the thermal energy ($\sim$ few $k_B T$) next to HOMO: i.e., the shallow trap energy is $0 < E_a \leq$ few $k_BT$, where $k_B$ is the Boltzmann's constant. Therefore, shallow traps remain partially depopulated by thermal excitations, regardless of the concentration of field-induced charges. This is contrary to the deep traps (i.e., the states with $E_a \gg k_BT$) that remain permanently filled with charges injected from the contacts at $|V_g| > |V_{th}|$. This important difference between deep and shallow traps helps explain the observed increase of $V_{th}$ with cooling [3,16].

The trap-free mobility, $\mu_0$, depends only on intrinsic factors, such as molecular packing, polaronic self-localization, dynamic disorder (phonons and thermal fluctuations of molecules), and competition between coherent and hopping motion of polarons [17,18,19,20]. The dynamic thermal disorder is especially large in van der Waals crystals [19], resulting in a very short momentum relaxation time ($\tau_{rel}$) and a small mean free path of polarons, $\lambda = \upsilon \cdot \tau_{rel}$, where $\upsilon = \sqrt{2k_BT/m^*}$ is the thermal velocity of non-degenerate polarons [21], and $m^*$ is their effective mass. Typically, $\lambda$ is of the order of one lattice constant and is much smaller than the average distance between the traps in the channel ($d'$): $\lambda \ll d' = 1/\sqrt{N} = 1/\sqrt{(N_0 + \Delta N)}$, where $N_0$ is the initial density of shallow traps, and $\Delta N$ is the density of extra (effective) shallow traps introduced in our experiment.

The small mean free path justifies the notion of diffusive motion of polarons between the shallow traps, and therefore the diffusion relationship can be applied:

$$\tau = d^2/D = \frac{1}{(N_0 + \Delta N)D}. \qquad (2)$$

Here, $d$ is the average distance between polaron's trapping sites which depends on the trapping cross section and, generally speaking, might differ from $d'$. However, $d$ in molecular crystals is of the same order of magnitude as $d'$ (see below), and the expression $1/(N_0 + \Delta N)$ can be used for $d^2$ in (2). The diffusion coefficient ($D$) for a non-degenerate system can be estimated according to Einstein's formula in two dimensions: $D = 4k_BT\mu_0/e$. It can be seen from eq. (1) that the reported effect, - a decrease of $\mu$ as a result of the photo-induced electron transfer, can only be observed in a trap-dominated regime. Indeed, in the intrinsic regime ($\tau \gg \tau_{tr}$), both numerator and denominator in (1) vary as $1/N$, and $\mu \cong \mu_0$ remains unaffected by changing $N$, unless the condition $\tau \gg \tau_{tr}$ is violated at large $\Delta N$. This suggests that the observation of $\mu$ decreasing with a photo-induced charge transfer can be used as evidence of a trap-dominated operation of an OFET; here, it shows that our tetracene devices are trap-dominated, and that $\mu_0$ could be higher than the typically observed effective mobilities: $\mu = 0.1 – 1.6$ cm$^2$/Vs [6,16,22].

Let us estimate $N_0$ - the density of shallow traps in the channel of as-prepared devices, using the experimental dependence $\mu(\Delta N)$. By substituting (2) into the expression (1) for the trap dominated regime, we obtain:

$$\mu(\Delta N) \approx \mu_0 \frac{\tau}{\tau_{tr}} = \left(\frac{\mu_0}{D\tau_{tr} N_0}\right) \frac{1}{1+\Delta N / N_0} = \mu^{(0)} \frac{1}{1+\Delta N / N_0}, \quad (3)$$

where $\mu^{(0)} \equiv \mu(\Delta N=0) = 0.5$ cm$^2$/Vs is the mobility of as-prepared device. The expression (3) has only one unknown parameter $N_0$, and by fitting $\mu(\Delta N)$ in the right part of Fig. 2 with (3), we find $N_0 = (3 \pm 0.5) \cdot 10^{11}$ cm$^{-2}$. Note that despite the fact that the maximum density of the field-induced carriers in our OFETs is few times higher than $N_0$, the transport in these devices is still trap dominated. This is because $N_0$ represents the fraction of shallow traps that are "*instantaneously* *empty*", and therefore $N_0$ could be considerably smaller than the total density of shallow states. Indeed, $d$ in (2) is the distance between the empty traps, in which a moving polaron can be trapped.

Let's now estimate the trapping time, $\tau_{tr}$. It can be seen from (3) that $\tau_{tr} = e/(4k_B T \cdot N_0 \cdot \mu^{(0)})$, which gives $\tau_{tr} \approx 50$ ps. It is useful, however, to derive a general formula for $\mu$ as a function of shallow trap density, $\mu(\Delta N)$, expressed in terms of $\tau_{tr}$ and $\Delta N$ and applicable for any $\tau/\tau_{tr}$. By substituting (2) into (1), and noticing that the effective mobility of as-prepared OFETs is $\mu^{(0)} \equiv \mu(\Delta N=0) = \mu_0/(1 + N_0 D \tau_{tr})$, we obtain for *diffusive motion of polarons in the presence of shallow traps*:

$$\mu(\Delta N) = \frac{\mu^{(0)}}{1 + (4k_B T / e) \cdot \mu^{(0)} \cdot \tau_{tr} \cdot \Delta N} \quad (4)$$

This formula contains only one parameter - the trapping time, $\tau_{tr}$. By fitting the experimental dependence $\mu(\Delta N)$ in the right part of Fig. 2 with (4), we obtain a similar value, $\tau_{tr} = 50 \pm 10$ ps. Note that no assumptions regarding the magnitude of $\tau_{tr}/\tau$ has been made to derive (4), and, therefore, this formula is applicable both to the trap-dominated and to the intrinsic regimes. Indeed, using $\mu \approx \mu_0 \cdot \tau/\tau_{tr}$, we will obtain the same formula (4). In the opposite limit of the intrinsic conduction ($\tau \gg \tau_{tr}$), the mobility of as-prepared devices is $\mu^{(0)} = \mu_0$, and (4) reduces to $\mu(\Delta N) = \mu_0/(1 + D \cdot \tau_{tr} \cdot \Delta N)$, where $D \cdot \tau_{tr} \cdot \Delta N = D \cdot \tau_{tr} / d^2 = \tau_{tr} / \tau \ll 1$, so that $\mu(\Delta N) = \mu_0$

= *const*. This limiting case of (4) is consistent with the observed *independence* of $\mu$ on the photo-induced charge transfer in OFETs operating in the intrinsic regime [13]. It should be noted that, given a certain energy distribution of traps, $\tau_{tr}$ is a phenomenological parameter that characterizes the entire ensemble of shallow traps.

Below, we argue that several assumptions made in our analysis can be justified. In eq. 2, it has been assumed that the average distance between polaron trapping sites is similar to the average distance between the shallow traps in the channel: $d \sim d' \equiv 1/\sqrt{N}$. This is not necessarily the case, and it depends on a "cross section" of polaron trapping, characterized by the "radius" of the trap ($r$). For two dimensional diffusive motion of polarons with a capture parameter $r$, the area "swept" during time $t$ can be expressed as: $s = 2r \cdot v \cdot (t - N^{(1)} \cdot \tau_{tr})$, where $N^{(1)}$ is the number of trapping events occurred on the trajectory, $N^{(1)} = N \cdot s$. Note also that $t = N^{(1)} \cdot (\tau_{tr} + \tau)$. Using this, we obtain the average time of diffusive motion of polarons between consecutive trapping events, $\tau$, in terms of the trap density, $N$, and the radius of capture, $r$: $\tau = \dfrac{1}{2rNv}$. Therefore, the ratio $d/d'$ is:

$$\frac{d}{d'} = \sqrt{D \cdot \tau \cdot N} = \sqrt{\frac{D}{2rv}} = \sqrt{\frac{m^* \cdot v \cdot \mu_0}{e \cdot r}} = \sqrt{\frac{\lambda}{r}} \ . \tag{5}$$

In (5), we used the relationship between the intrinsic mobility, momentum relaxation time and the effective mass: $\mu_0 = \dfrac{e\tau_{rel}}{m^*} = \dfrac{e\lambda}{m^* \cdot v}$. The effective "size" of traps in organic molecular crystals, $r$, is of the order of one lattice constant [1], as well as the mean free path $\lambda$, and, therefore, $d \approx d'$.

We assumed that each transferred electron creates one shallow trap in the channel. This assumption is reasonable considering the low densities of the transferred charge $\Delta N < 3 \cdot 10^{11}$ cm$^{-2}$ (Fig. 2), resulting in dilute static electrons near the interface. In addition, Fig. 2 shows that the density of deep traps, $N_{deep}$, remains nearly constant, suggesting that the electrons immobilized in the insulator do not create deep traps and, therefore, contribute only to the shallow states.

It was also assumed that generation of new shallow traps only increases $N$, while the trapping time, $\tau_{tr}$, remains unaffected, thus leaving the characteristic energy of the whole ensemble of traps, $E_a$, unchanged. This assumption can be verified by performing $\mu(T)$ measurements before and after the photo-induced electron transfer. Fig. 3 shows $\mu(T)$ dependences for a device in the two states with different densities of shallow traps: $\Delta N = 0$ and $\Delta N = 2.5 \cdot 10^{11}$ cm$^{-2}$. It can be seen that the temperature dependence of the mobility has the form: $\mu \propto \gamma(N) \cdot \varphi(T)$, i.e., despite the fact that the mobility is significantly reduced (note the semi-log scale), the temperature dependent factor, $\varphi(T)$, remains the same, suggesting that the characteristic energy of traps $E_a$ and, therefore, $\tau_{tr}$ have not been affected by changing the density of shallow traps. In order to obtain an estimate of the energy scale of the trapping process, we show the data in an Arrhenius form (the inset in Fig. 3), indicating that both dependences have the same activation energy $E_a \approx 20$ meV, comparable to $k_BT$ in this temperature range. Here, we have neglected the temperature dependence of the intrinsic mobility $\mu_0(T)$; this however does not change the conclusion of independence of $E_a$ on $N$.

These observations are consistent with our understanding of shallow traps. Indeed, at sufficiently large $V_g$, only thermal excitations determine the range of in-gap states near the band that can be de-populated. Therefore, the characteristic energy of these states must always be of the order of $k_BT$, which is consistent with $E_a$ typically observed in contact-corrected devices [16]. If the generated new traps had energy outside of the few-$k_BT$ window, they would become deep traps and would contribute only to the field-effect threshold; if they have $E_a$ within ~ $k_BT$ from HOMO, they automatically contribute to the density of shallow traps $N$, without affecting their energetics. This does not imply, however, that these "photo-induced" traps have similar microscopic origin to those in pristine devices.

To conclude, we report a monotonic decrease of the field-effect mobility as a result of the photo-induced electron transfer from the semiconductor into the dielectric in OFETs. The effect is non-volatile and can only be observed in trap-dominated devices. Our interpretation is based on the generation of extra shallow traps in the channel as a result of the Coulomb attraction of polarons to the electrons immobilized in the dielectric, in a vicinity of the conduction channel. Note that independence of $\mu$ on the hole transfer is consistent with the model of diffusive transport limited by a strong thermal disorder [19], for which adding few more scattering centers can not alter $\mu$. The effect can be used to measure the dependence of $\mu$ on the density of shallow traps and to extract the (initial) density of shallow traps and average trapping time. It is important to emphasize the purely electronic nature of our measurements providing a characterization of *shallow traps* strictly relevant to field-effect conduction at the semiconductor/dielectric interface.

This work has been supported by the NSF grants DMR-0405208 and ECS-0437932. We are indebted to Prof. M. E. Gershenson for helpful scientific discussions and significant technical support that made this work possible.

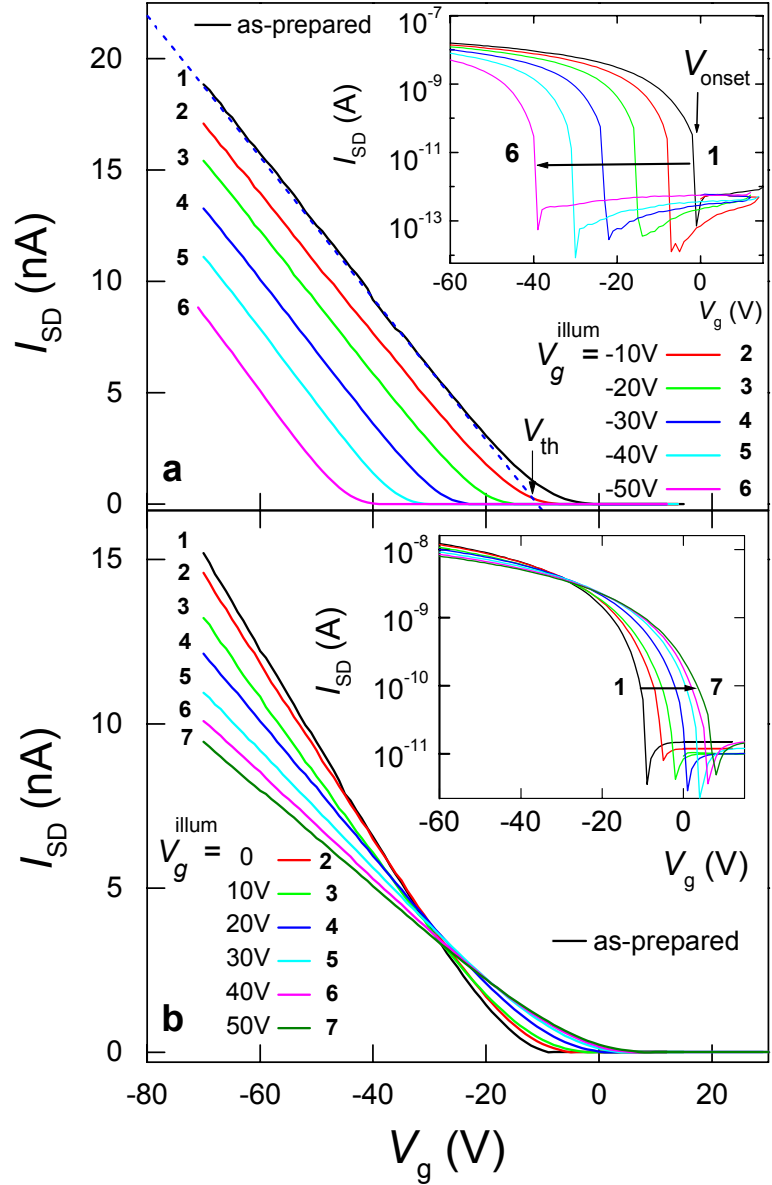

**Fig. 1.** $I_{SD}(V_g)$, of tetracene single-crystal OFET measured after the photo-induced transfer of holes (**a**) and electrons (**b**) from tetracene into parylene ($V_{SD}$ = 5 V, $T$ = 300 K, $V_g$ is swept from positive to negative). The charge transfer is achieved by application of a gate voltage $V_g^{illum}$ under illumination of the OFET's channel with a white light. Each curve has been recorded in the dark, following the corresponding "gating under illumination" treatment. The insets show the same data on a semi-log scale. The rate of $V_g$ sweep was 0.33 V/s, corresponding to ~ 5 min recording time per curve.

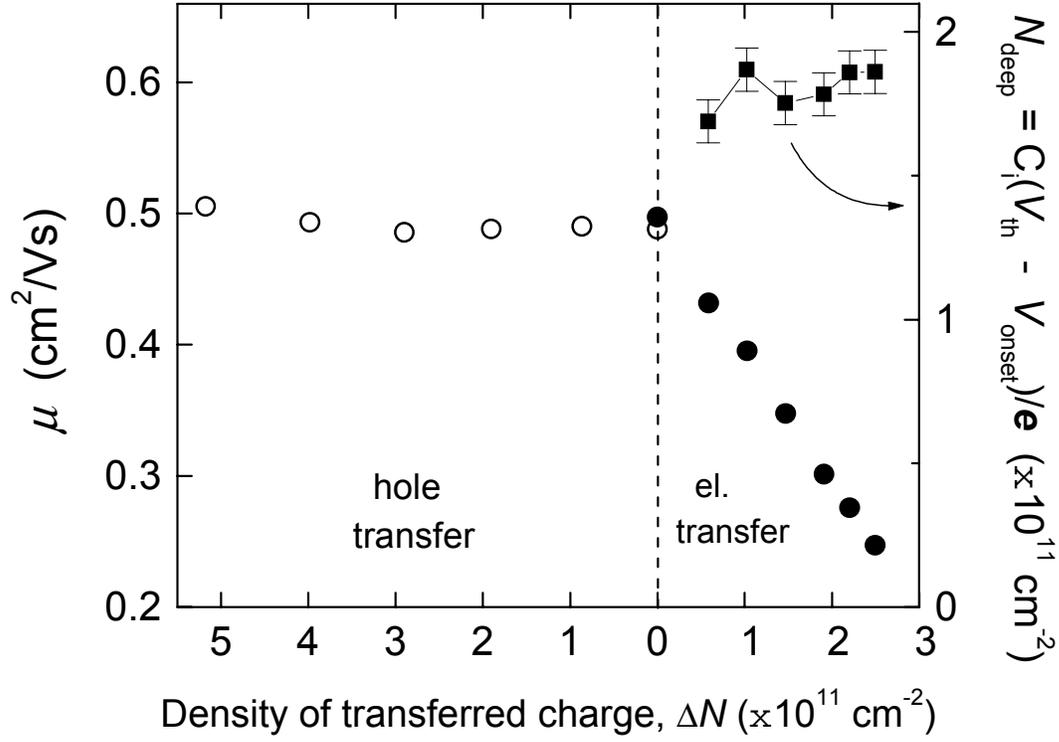

**Fig. 2.** The field-effect mobility $\mu$ (circles, left axis) and the density of deep traps (squares, right axis) as a function of the density of transferred charge, $\Delta N$. The values of $\mu$ have been determined from the linear portions of $I_{SD}(V_g)$ curves shown in Fig. 1 using the conventional field-effect equations (see, e.g., [2]). $N_{deep}$ is determined using $V_{th}$ and $V_{onset}$ from Fig. 1(b).

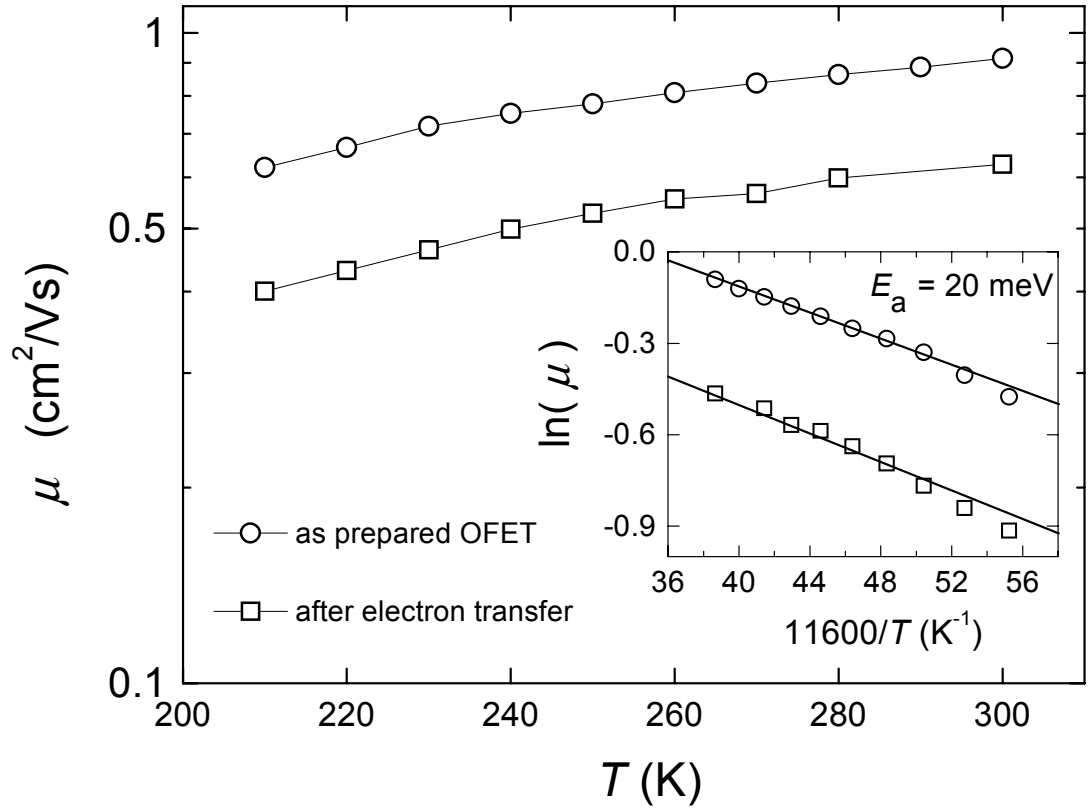

**Fig. 3.** Temperature dependence of the mobility $\mu(T)$ of an OFET before (○) and after (□) the increase of the shallow trap density in the channel by $2.5 \cdot 10^{11}$ traps/cm$^2$. The inset shows an Arrhenius plot of the data. The activation energy in both cases is $E_a = 20$ meV, which is of the order of the thermal energy, $k_BT$, in this temperature range.